\documentstyle[prb,aps]{revtex} 

\begin{document} 
\preprint{NSF-ITP-99-119}
\draft

\title{\Large{Zener transitions between dissipative Bloch bands.  II: Current
Response at Finite Temperature}}

\author{ Xian-Geng Zhao }

\address{CCAST (World Laboratory) P.O.  Box 8730, Beijing 100080, China\\
      Institute of Applied Physics and Computational Mathematics, \\P.O.
      Box 8009, Beijing 100088, China\\}

\author{Daniel W. Hone}

\address{Institute for Theoretical Physics, University of California, Santa
Barbara, CA 93106 }

\date{\today} \maketitle

\begin{abstract}
We extend, to include the effects of finite temperature, our earlier 
study of the interband dynamics of electrons with Markoffian dephasing
under the influence of uniform static electric fields.  We use a simple
two-band tight-binding model and study the electric current response 
as a function of field strength and the model parameters.  In addition to
the Esaki-Tsu peak, near where the Bloch frequency equals the damping
rate, we find current peaks near the Zener resonances, at equally spaced
values of the inverse electric field.  These become more
prominenent and numerous with increasing bandwidth (in units of the
temperature, with other parameters fixed).  As expected, they broaden
with increasing damping (dephasing).

\end{abstract}

\pacs{PACS numbers:  73.20.Dx, 71.70.Ej, 73.40.Gk}

\widetext

\section{Introduction} 

In a previous paper\cite{yanzhao} (henceforth
referred to as I) we began a study of the effects on interband transitions
of the scattering of electrons from static imperfections in a semiconductor
superlattice.  A uniform static electric field was applied.  It has long been
recognized that scattering destroys the coherence necessary to sustain the
Bloch oscillations predicted in such a field, and in practice this delayed
the observation of these oscillations until the development of semiconductor
superlattices\cite{esaki,expt}.  But the scattering destroys other types of
interesting coherent motion, as well.  Dunlap and Kenkre\cite
{dunken,dunkenpla}, in particular, looked at the effects on dynamic
localization of electrons in time periodic electric fields.  We were
interested in I in a multiband effect, the Rabi 
oscillations\cite{xgniu2,rjs,rjs2} of electron population between bands 
of a crystal in a uniform static
electric field of appropriate magnitude --- near an avoided crossing of the
two interpenetrating Wannier-Stark Ladders (WSL) arising essentially from
different bands\cite{rjs}.  These are the so-called Zener resonances.
Within a simple two-band tight-binding model we demonstrated the destruction
of localization (of occupied electron states) to a single miniband by the 
dephasing associated with the scattering.
The decay rate for the approach to steady state band populations exhibits
sharp peaks at values of the static electric field which give Zener
resonances in the absence of scattering.  The specific approximations in I,
however, effectively limited the results to the case of infinite
temperature.  In particular, the steady state was assumed to be equal
population of the two bands.  In this paper we remove that restriction, to
discuss these effects at finite temperature.  The necessary modification
was described in Ref. (\onlinecite{dunken}), namely relaxation of band
populations toward values set by the Boltzmann factors describing thermal
equilibrium.  This will allow us, in particular, to look at the electric
current, a quantity of obvious experimental interest which vanishes in the
infinite temperature limit.

\section{Model}

We consider the same model that we treated in I, a standard simple
tight-binding model\cite{fukuyama} of a two-band system in a static electric
field E.  The Hamiltonian\ can be written as 
\begin{eqnarray} H =
&&\sum_n\bigg[(\Delta_a + n\omega_B)a_n^\dagger a_n + (\Delta_b +
n\omega_B)b_n^\dagger b_n \nonumber\\ &&- (W_a/4)(a^\dagger_{n+1} a_n +
h.c.)  + (W_b/4)(b^\dagger_{n+1} b_n + h.c.)\nonumber\\ && + eER(a^\dagger_n
b_n + b^\dagger_n a_n)\bigg].  
\label{HAM} 
\end{eqnarray} 
Here the
subscripts label the lattice sites and the lower and upper minibands are
designated by symbols $a$ and $b$, respectively.  We have introduced the
notation $\omega_B\equiv eEd$, where $d$ is
the lattice constant, for the Bloch frequency, which will appear
often below.  The first two terms describe the site energies of the Wannier
states in the presence of the electric field, and $W_{a,b}$ are the widths
of the isolated ($ E=0 $) minibands induced by nearest neighbor hopping:
$\epsilon_{a,b}(k) = \Delta_{a,b} \mp (W_{a,b}/2) \cos k $, where the 
dimensionless wave vector $k$ is in units of the inverse lattice constant $d$.  
The last term is the on-site electric dipole coupling
between minibands; $eR$ is the corresponding dipole moment.  This
Hamiltonian does neglect Coulomb interactions and electric dipole elements
between Wannier states on different sites, but it contains the essential
physics for the problem\cite{rjs,rjs2,fukuyama,honexg}.  Note that the
hopping parameters $W_{a,b}$ are written here with opposite signs, so that
with both parameters positive the band structure at $E=0$ is of the standard
nearly free electron character, with direct band gaps at the zone boundary.
But the calculation to follow is valid for arbitrary signs of the
parameters.

 It is easily shown\cite{fukuyama,honexg} that the exact spectrum of $H$ is
two interpenetrating Wannier-Stark Ladders.  But what do the corresponding
states represent in terms of the occupation of the original bands as a
function of time, and what is the influence of scattering?  For vanishing
dipole matrix element between bands, $R=0$, there is no interband mixing.
Each of the two bands gives rise to a single WSL.  Clearly, when the
electric field amplitude is such that the ladders become degenerate, even
small values of $R$ lead to strong interband mixing.  The crossing of the
ladders is ``avoided" by any finite $R$, and the behavior near those avoided
crossings (the Zener resonances) is of particular importance and interest.
In general there are peaks in the current response at those values of the
electric field, but the peaks are broadened by increasing temperature, as
well as by decreasing bandwidth relative to band separation.

We start by defining the density matrix in the representation of the two
bands,
\begin{equation}
\rho(t)=\sum_{ijmn}\rho_{mn}^{ij}\xi_{m}^{i\dagger}\xi_{n}^{j},
\end{equation} 
where $i,j = 1$ or $2$ are band indices:
$\xi_{m}^{1\dagger}$ ($\xi_{m}^{1}$) and $\xi_{m}^{2\dagger}$
($\xi_{m}^{2}$) designate $a_m^{\dagger}$ ($a_m$) and $b_m^\dagger$ ($b_m$),
respectively.  Since we are interested in the dynamics of occupation of
various band states, it is convenient to work in a wave vector basis, by
Fourier transforming the density matrix.  In general, since $\rho_{mn}$ is
not translationally invariant (a function only of $m-n$), we have a full set
$\rho^{ij}_{kq} = \sum_{mn} \rho_{mn}^{ij}(t)\exp[-ikm+iqn]$ of Fourier
components.  But we will be interested in the wave vector diagonal band
occupation numbers $\rho^{ij}_{kk}(t) \equiv \rho^{ij}(k,t)$.  Then at
finite temperature $T$ we insist that the wave vector and band diagonal
occupation number relax to the thermal equilibrium value, 
\begin{equation}
\rho^{ii}_{kk} \rightarrow \rho^i_T(k) \equiv e^{-\beta\epsilon_i(k)} \Big/
\sum e^{-\beta\epsilon_j(q)}, \label{rhot} 
\end{equation} 
where $\beta = 1/k_BT$, the sum in
the partition function is over $j=1,2$ and over all wave vectors $q$, and
the band energies are those given above, 
\begin{equation} 
\epsilon_{1,2}(k) = \Delta _{a,b} \mp (W_{a,b}/2)\cos k ~.
\end{equation}

Within a constant relaxation rate approximation\cite{dunken}, the density
matrix $\rho(k,t)$ satisfies the following stochastic Liouville equation
(SLE) (we set $\hbar=1$ throughout this paper), 
\begin{equation}
i\frac{d\rho}{dt}=[H, \rho(t)]-i\Gamma \left[\rho(t) - \rho_T\right].
\label{SLE} 
\end{equation} 
Here each of $H$, $\rho$, and $\rho_T$ is
labelled by (the same) wave vector $k$.  The operator $\Gamma $ describes
the relaxation of the off-diagonal elements of $\rho$ through dephasing:
\begin{equation} 
\Gamma [\rho - \rho_T] = \sum_{ij}\alpha_{ij}\left[ \rho^{ij}(k,t) 
 -\delta_{ij} \rho_T(k) \right] \xi^{i\dagger}(k)\xi^j(k).  
\end{equation} 
The utility of this simplest form
of the SLE has been discussed by Kenkre and collaborators (see Ref.
\onlinecite{dunken} and references therein).  The parameters $\alpha_{ij}$
measure the loss of phase coherence between sites, or the scattering
lifetime of band states labeled by quasimomentum.

As in I it is convenient to introduce the linear combinations of density
matrix elements:  
\begin{eqnarray}
\rho_+(k,t)=\rho^{11}(k,t)+\rho^{22}(k,t)~,\\
\rho_-(k,t)=\rho^{11}(k,t)-\rho^{22}(k,t)~,\\
\rho_{+-}(k,t)=\rho^{12}(k,t)+\rho^{21}(k,t)~,\\
\rho_{-+}(k,t)=i[\rho^{21}(k,t)-\rho^{12}(k,t)]~.  
\end{eqnarray} 
For simplicity, we also take $\alpha_{11}=\alpha_{22}=
\alpha_{12}=\alpha_{21}=\alpha$ to reduce the number of parameters in the
theory.  Then the SLE (\ref{SLE}) has the explicit components
\begin{equation} 
\frac{\partial}{\partial
t}\rho_+(k,t)-\omega_B\frac{\partial}{\partial k} \rho_+(k,t)=
- -\alpha\left[\rho_+(k,t)-\rho^+_T(k)\right]~,
\label{rplus} 
\end{equation}
\begin{equation} 
\frac{\partial}{\partial
t}\rho_-(k,t)-\omega_B\frac{\partial}{\partial k} \rho_-(k,t)= -2eER
\rho_{-+}(k,t)-\alpha\left[\rho_-(k,t)-\rho^-_T(k)\right]~, 
\end{equation} 
\begin{equation} 
\frac{\partial}{\partial
t}\rho_{+-}(k,t)-\omega_B\frac{\partial} {\partial k} \rho_{+-}(k,t)=
\Bigl(\Delta - W\cos k\Bigr) \rho_{-+}(k,t) -\alpha\rho_{+-}(k,t)~,
\end{equation} 
\begin{equation} 
\frac{\partial}{\partial
t}\rho_{-+}(k,t)-\omega_B\frac{\partial} {\partial k} \rho_{-+}(k,t)=
- -\Bigl(\Delta - W\cos k\Bigr) \rho_{+-}(k,t) +2eER
\rho_-(k,t)-\alpha\rho_{-+}(k,t)~.  
\end{equation} 
Here we have used the
simplified notation $\Delta\equiv\Delta_a - \Delta_b$ and $W \equiv
(W_a+W_b)/2$.  The equation (\ref{rplus}) for $\rho_+$ is decoupled from
the others, and is readily integrated to give 
\begin{equation} 
\rho_+(k,t) =
e^{-\alpha t}\left\{\rho^+_T(k+\omega_Bt) + \alpha\int^t_0 dt'\, e^{\alpha
t'} \rho^+_T[k+\omega_B(t-t')]\right\}.
\label{rhoplus}   
\end{equation} 
The equations for
$\rho_-(k,t),~~\rho_{+-}(k,t),$ and $\rho_{-+}(k,t)$ can be reduced to the
following ordinary differential equations in an accelerated
basis\cite{kria}, $k(t) = k-\omega_B t$ or, equivalently, in the transverse
or vector gauge discussed in Ref.\ \onlinecite{honexg}, 
\begin{equation}
\frac{d}{dt}X(k,t)=-2eER~Z(k,t)-\alpha[X(k,t)-\rho^-_T(k-\omega_Bt)],
\end{equation} 
\begin{equation} 
\frac{d}{dt}Y(k,t)= [\Delta -
W\cos(k-\omega_B t)]Z(k,t) -\alpha Y(k,t), 
\end{equation} 
\begin{equation}
\frac{d}{dt}Z(k,t)= -[\Delta - W\cos(k-\omega_B t)]Y(k,t)
+2eER~X(k,t)-\alpha Z(k,t)~.  
\end{equation}

Here $X(k,t)=\rho_-(k-\omega_B t,t)$, $Y(k,t)=\rho_{+-}(k-\omega_B t,t)$,
and $Z(k,t)=\rho_{-+}(k-\omega_B t,t)$.  The structure is exactly the
same as Eqs.  (17 - 19) in I, except for the $k$-dependent relaxation of $X$
here.  As we did there, we can integrate these equations analytically as a
perturbation series in the parameter $\mu\equiv 2eER$, which characterizes
the electric dipole coupling between bands.  To lowest nontrivial (second)
order in $\mu$ we find 

\begin{eqnarray} 
\rho_-(k,t) = e^{-\alpha t} \bigg\{\rho^-_T(k+\omega_Bt)
&+& \alpha \int_0^t dt'\,e^{\alpha t'}\rho^-_T[k+\omega_B(t-t')] -\mu^2
\int_0^t dt'\int_0^{t'} dt''\, \Big[\rho^-_T(k+\omega_Bt)\nonumber\\ 
&+&\alpha \int_0^{t''} d\tau\,e^{\alpha \tau}\rho^-_T[k+\omega_B(t-\tau)]
\bigg] \cos \int_{t''}^{t'} d\tau '\left[\Delta
- -W\cos\left(k+\omega_B(t-\tau ') \right)\right]\Big\}~.
\label{rhominus} 
\end{eqnarray}

\section{Current}

We turn now to the calculation of the interesting physical quantity, the
current $j(t)$ along the superlattice direction.  In the band $i=a,b$ the 
instantaneous current is given by the sum over wave vectors of the relevant 
electron velocity $v_i(k) = (W_i d/2)\sin k$ times the number density 
in that band at that wave vector, $n_0\rho_i(k,t)$, where $n_0$ is the 
number of carriers per unit area in each cell of the superlattice.  In terms 
of the convenient quantities $\rho_{\pm}(k,t)$ we then have 
\begin{equation} 
j(t) = \int_0^{2\pi}\frac{dk}{8\pi}\left[(W_a-W_b)\rho_+(k,t) +
2W\rho_-(k,t)\right] n_0 d\sin k.  
\end{equation} 
Of particular interest is the
steady state long time average of this, 
\begin{equation} 
\langle j\rangle = \lim_{T_0\rightarrow\infty} \frac{1}{T_0}
\int_0^{T_0} dt \, j(t).  
\label{jav}
\end{equation} 
To second order in $\mu$ we have the probability densities
$\rho_{\pm}(k,t)$, in Eqs.  (\ref{rhoplus}) and (\ref{rhominus}), 
and we find
\begin{eqnarray} 
\frac{\langle j\rangle}{n_0 d} &=& \left(\frac{W_a}{2}\right)\left(\frac
{\alpha\omega_B}{\alpha^2+\omega_B^2}\right)[C_1+(W_b/W_a)C_2] \nonumber\\
&-& \left(\frac{W}{4}\right)\left(\frac{\alpha\mu}{\alpha^2+
\omega_B^2}\right)^2 [C_1+C_2] \sum_{\ell=-\infty}^{\infty} \left[
D_{\ell}^- + D_{\ell}^+\right]J_{\ell}^2(W/\omega_B) , 
\label{javg}
\end{eqnarray}
with 
\begin{equation} 
C_1 \equiv
\frac{e^{-\beta\Delta}I_1(\beta W_a/2)} {e^{-\beta\Delta}I_0(\beta
W_a/2)+ I_0(\beta W_b/2)} ~,
\end{equation} 
\begin{equation} 
C_2 \equiv \frac{I_1(\beta W_b/2)} {e^{-\beta\Delta}I_0(\beta W_a/2)
+ I_0(\beta W_b/2)}~, 
\end{equation} 
\begin{equation} 
D_{\ell}^{\pm} \equiv \frac{\alpha^2-\omega_B^2-2\omega_B
[(\ell+1)\omega_B\pm\Delta]}{\alpha^2+[(\ell+1)\omega_B\pm\Delta]^2}~,
\end{equation}
where $I_n$ is the modified Bessel function and $J_n$ the
ordinary Bessel function of order $n$. 
The interband effects are all contained in the second term on
the right hand side of (\ref{javg}), proportional to $\mu^2$.  The Zener
resonances, near $\Delta = n\omega_B$, with $n$ an integer, exhibit
themselves as peaks in the factors $D_{\ell}^-$.

We will look at the nonlinear conductance, the current as a function of
increasing electric field, which is conveniently parameterized in
the dimensionless form $\omega_B/\Delta$.
At sufficiently high fields, as the Wannier-Stark functions become increasingly
localized in space, the current falls off inversely with
$\omega_B$.  To the extent that the interband contribution can be neglected
(small $\mu$) the field dependence of the response is given by the factor
$\omega_B/(\alpha^2+\omega_B^2)$, with an ``Esaki-Tsu" peak\cite{esaki} near
$\omega_B = \alpha$.

There are two sources of temperature dependence in (\ref{javg}).  The first
is the explicit appearance of $\beta \equiv 1/k_BT$ in the factors $C_1$ and
$C_2$.  The other is implicit; the relaxation rate $\alpha$ is ordinarily
also temperature dependent.  Though that can be modeled for a specific
relaxation mechanism, we simply take it to be constant below.  At high
temperatures, with $\beta\Delta$, $\beta W_i \ll 1$, we have $C_1$ and $C_2$
approximately linear in $\beta$; the current $\langle j\rangle$ also
therefore falls off as $1/T$.  This is the well known effect of approaching 
uniform thermal population of the states throughout each band as the 
temperature rises; at complete uniformity the current vanishes.

All of these effects are seen in Fig.  1, where the interband matrix element
$\mu$ has been set equal to zero, and $\alpha/\Delta=0.1$. The second term
on the right hand side of Eq. (\ref{javg}) is then absent, and the
single remaining peak arises from the first term in that equation. To avoid a
multiplicity of parameters we have taken the two bandwidths to be equal,
$W_a=W_b=W = 0.8\Delta$.  In the remaining figures we have set the interband
coupling to be $\mu/\Delta = 0.4$.  Again $W_a=W_b=W$, with increasing
values $W/\Delta =$ 0.4 and 0.8 in Figs.  2  and 3,
respectively.  Each figure gives results for three values of the parameter
$\beta\Delta$, namely 1, 10 and 100.  The value of unity is representative
of a typical superlattice, of lattice parameter of order 100 \AA ~ and carrier
effective mass $m^*/m\approx 0.1$, at room temperature.  From Eq.  
(\ref{javg}) we see that the height in $\langle j\rangle$ of the 
$n$th peak (for values of $n$ up to 5 or 6, where
$\alpha^2\ll\omega_B^2$), at $\omega_B=\Delta/n$ is  approximately
proportional to $[n/(1+.01n^2)]^2J_{n-1}^2(nW/\Delta)$ .  The Bessel function
overlap factors increase rapidly with increasing bandwidth $W$ over the 
range we have chosen for $W$, and so we see more
well-defined Zener resonance peaks with increasing bandwidth , going from
Fig.  2 to Fig.  3.  The major effect of increasing the damping rate $\alpha$
is to broaden the peaks in the current.  Thus, for the particular choice
of $W/\Delta =$ 0.4 and $\mu/\Delta = 0.4$, we can compare the results for
$\alpha/\Delta = 0.05$ in Fig. 4 and $\alpha/\Delta = 0.2$ in Fig. 5 with 
the previous curves for $\alpha/\Delta = 0.1$ in Fig. 2.

\section{Conclusions}

Within a simple two-band tight binding model we have studied the 
electric current
response of a semiconductor superlattice subject to a finite uniform
electric field.  Relaxation processes have been assumed Markoffian, and
they have been described by a single parameter characterizing the rate
at which a nonequilibrium density is restored to thermal equilibrium,
using a stochastic Liouville equation.  We have been particularly 
interested in the interband transitions, which lead to current peaks
near the values of external field where Zener resonances occur.  We have
exhibited the variation in height and width of these peaks, as well as
in the overall current magnitude with temperature and with the various
parameters of the theory, including bandwidth, band separation,
interband dipole coupling, and relaxation rate.

The results we have obtained are perturbational in $\mu$; we do require
weak interband coupling. Moreover, we have limited the discussion to a
single pair of bands, assuming the impact of all other bands to be
negligible on the Zener tunneling between these two.  This can be 
realized in practice by using a dimerized semiconductor superlattice
with, for example, uniform  wells but alternating thick and thin
barriers between them.  With weak alternation of barrier thickness one
can adjust a pair of bands resulting from the doubling of the unit cell
to be well isolated from all other bands, and the predictions of this
paper can be studied in such a system by varying the external uniform
electric field.  Another way to observe the effects predicted here is to use
ultracold atoms in accelerating optical potentials. Recently,  
Rabi oscillations were observed in such a system, where the interband
coupling was generated by a small phase modulation \cite{Raizen}. Very 
low temperatures can be realized in these systems, and it should be
possible to observe the resonant structures in current that we have 
predicted in this paper.

This work was supported in part by the National Natural Science Foundations
of China under Grant No. 19725417, and in part by the U.S. National Science
Foundation Grant No. PHY94-07194.

\begin{figure}
\caption{Time averaged current in the absence of interband transitions 
($\mu=0$). The ratio of relaxation rate to band center separation is 
$\alpha/\Delta = 0.1$.  The bandwidths are $W_a=W_b=W=0.8\Delta $. }
\end{figure}

\begin{figure}
\caption{Time averaged current, as in Fig. 1, but with 
$\mu = W = 0.4\Delta$. }
\end{figure}
 
\begin{figure}
\caption{Time averaged current, as in Fig. 2, but with $W = 0.8\Delta$.}
\end{figure}

\begin{figure}
\caption{Time averaged current, as in Fig. 2, but with 
$\alpha = 0.05\Delta$.}
\end{figure}

\begin{figure}
\caption{Time averaged current, as in Fig. 2, but with 
$\alpha = 0.2\Delta$.}
\end{figure}

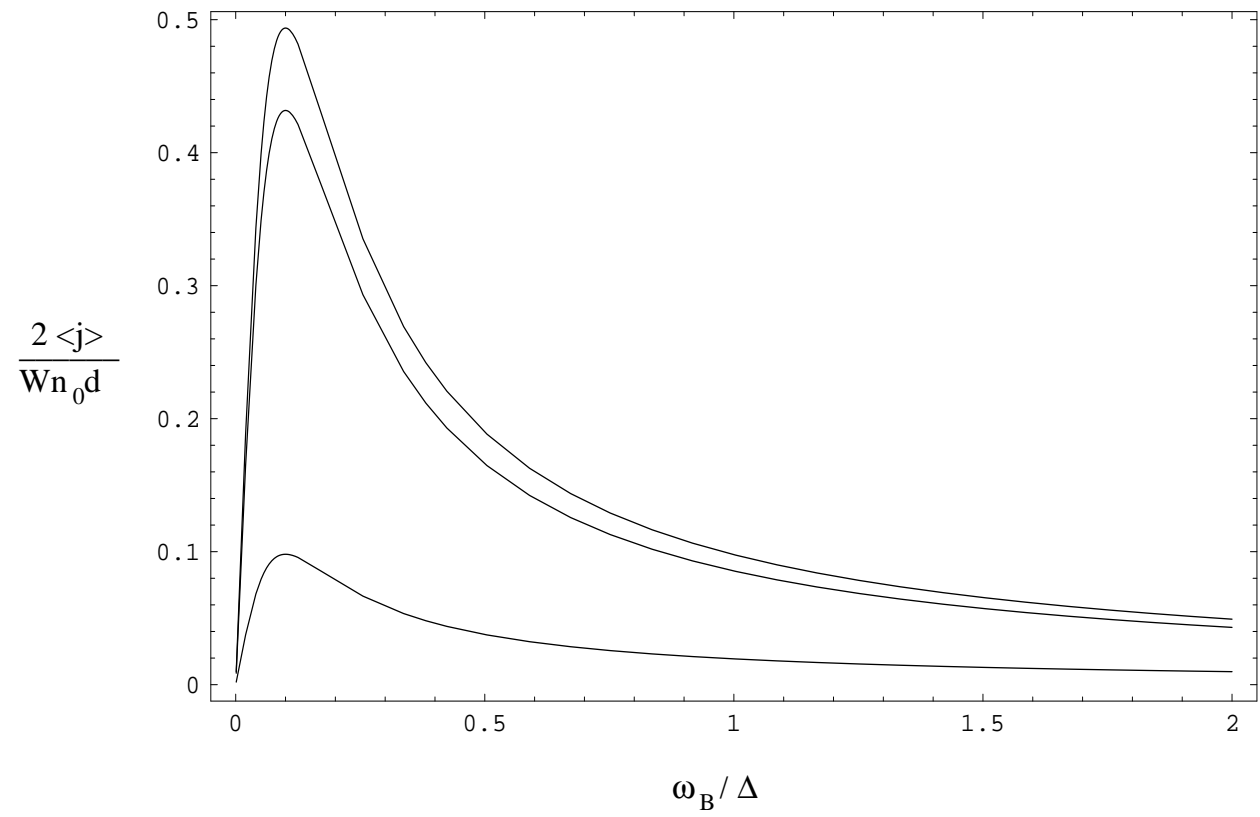
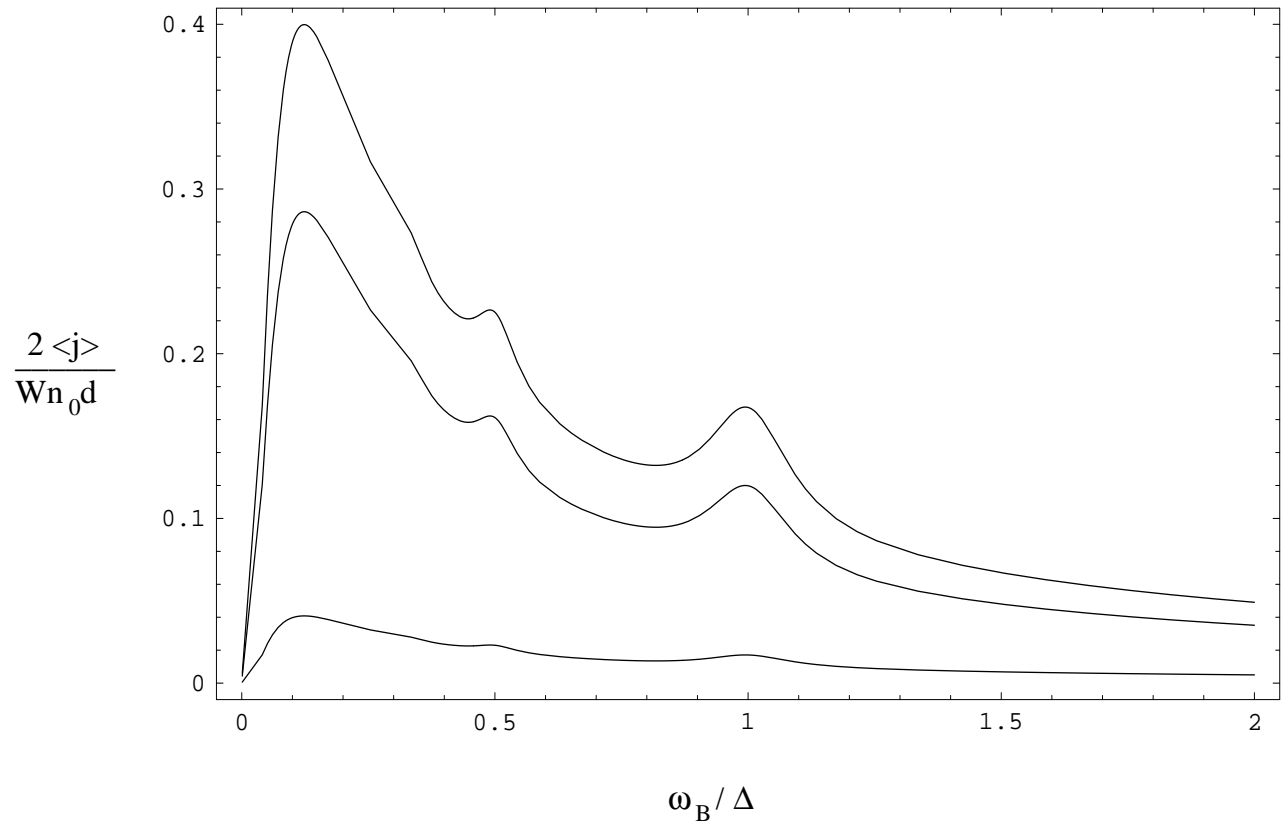
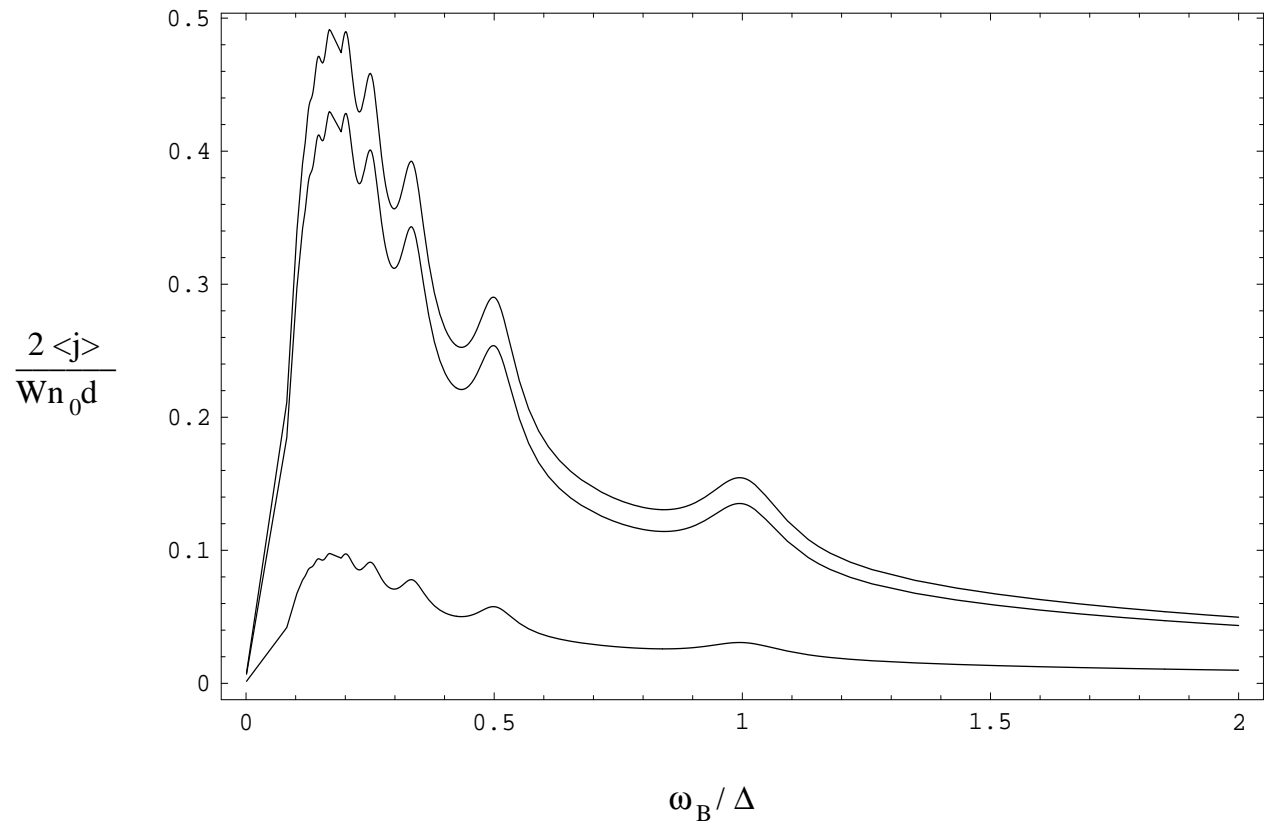
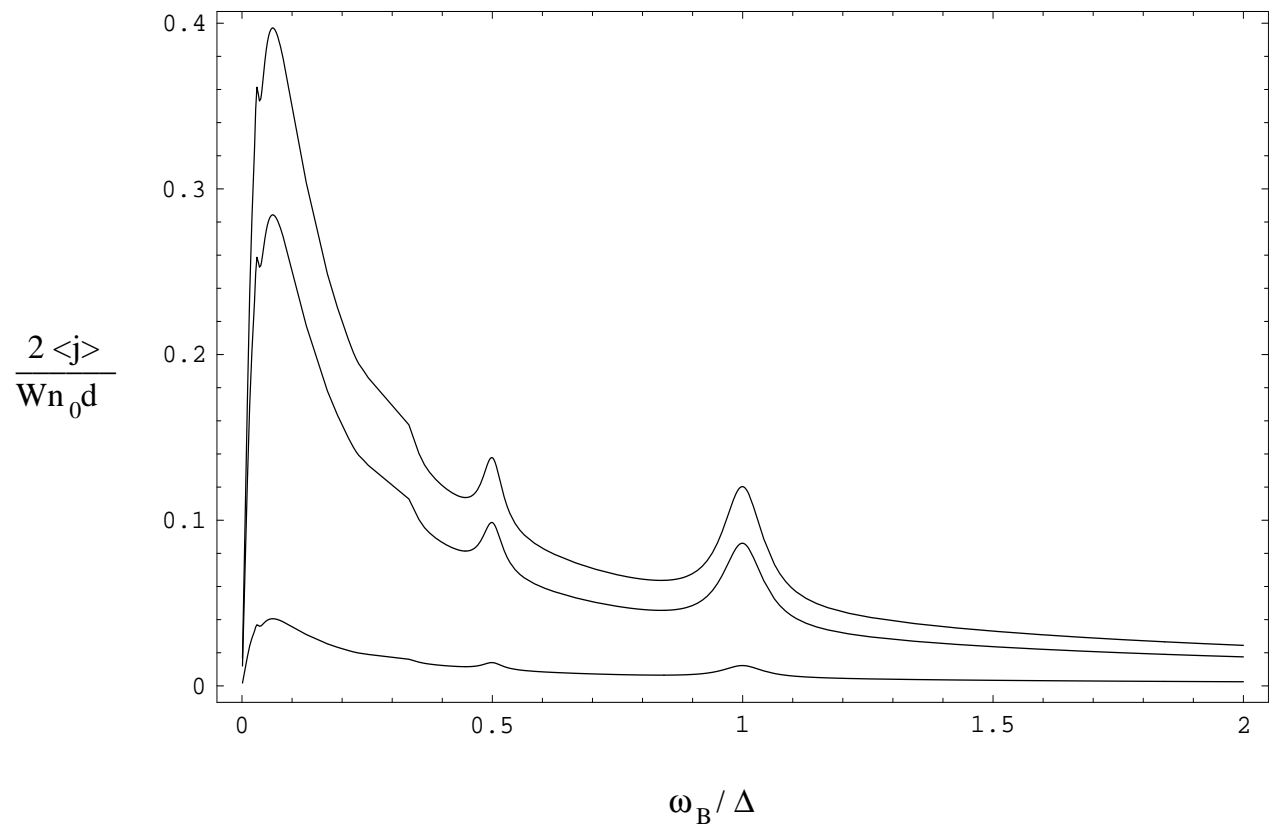
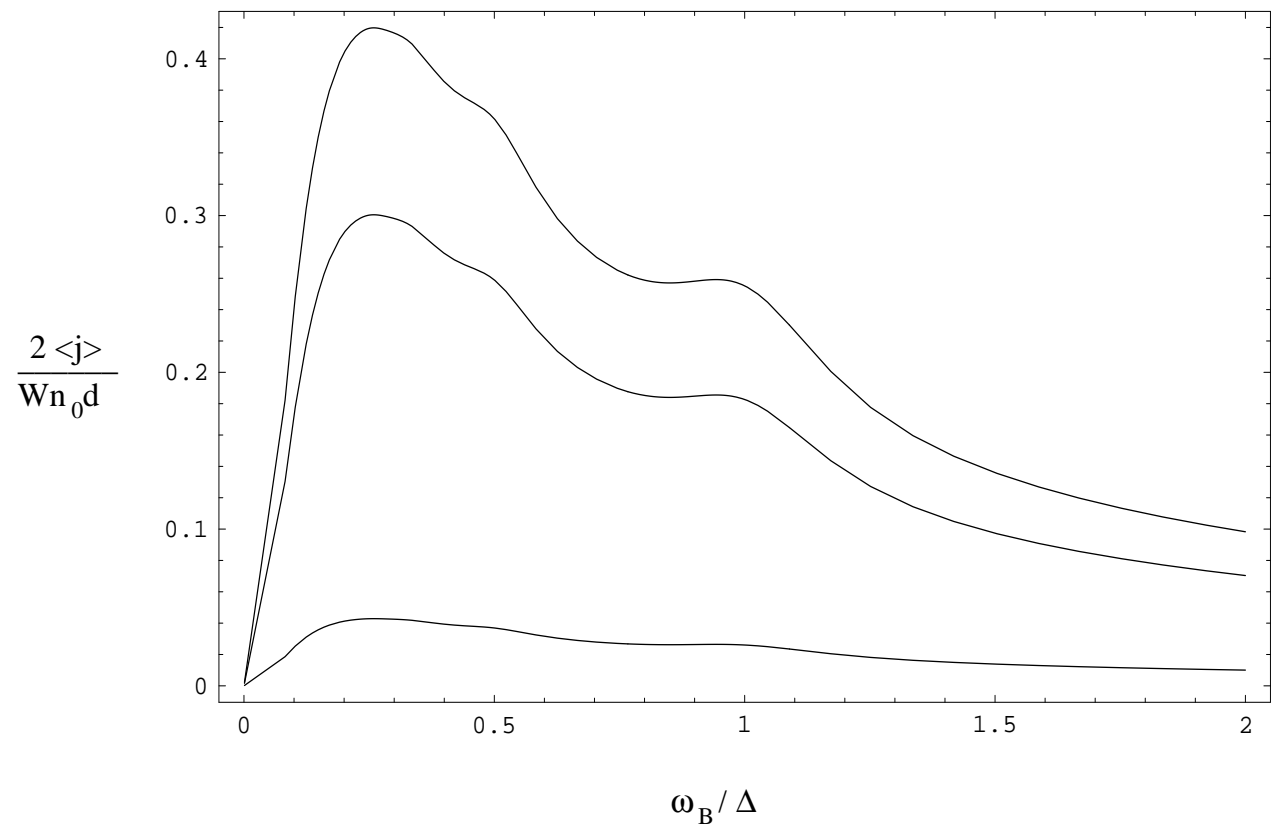
\end{document}